\documentclass[11pt]{article}
\usepackage{amsfonts,amssymb,amsmath}

\usepackage[english]{babel}
\usepackage{color}

\newtheorem{theorem}{Theorem}
\newtheorem{definition}{Definition}
\newtheorem{remark}{Remark}
\newtheorem{proof}{Proof}

\newtheorem{proposition}{Proposition}
\newtheorem{lemma}{Lemma}

\newcommand{\beq}{\begin{eqnarray}}
\newcommand{\eeq}{\end{eqnarray}}

\newcommand{\beqt}{\begin{eqnarray*}}
\newcommand{\eeqt}{\end{eqnarray*}}

\newcommand{\be}{\begin{equation}}
\newcommand{\ee}{\end{equation}}

\newcommand{\bl}{\begin{lemma}}
\newcommand{\el}{\end{lemma}}

\newcommand{\bt}{\begin{theorem}}
\newcommand{\et}{\end{theorem}}

\newcommand{\bd}{\begin{definition}}
\newcommand{\ed}{\end{definition}}

\newcommand{\bp}{\begin{proposition}}
\newcommand{\ep}{\end{proposition}}

\newcommand{\bpr}{\begin{proof}}
\newcommand{\epr}{\end{proof}}

\newcommand{\bi}{\begin{itemize}}
\newcommand{\ei}{\end{itemize}}

\newcommand{\ben}{\begin{enumerate}}
\newcommand{\een}{\end{enumerate}}

\newcommand{\Z}{\mathbb Z}

\newcommand{\E}{\mathbb E}

\newcommand{\s}{\ensuremath{\mathcal{S}}}

\newcommand{\om}{\ensuremath{\omega}}
\newcommand{\Om}{\ensuremath{\Omega}}

\newcommand{\La}{\ensuremath{\Lambda}}
\newcommand{\si}{\ensuremath{\sigma}}

\begin{document}

\title{{\bf One-sided versus two-sided stochastic descriptions.}}
 
\author{ Aernout C.D.  van Enter \footnote{ Bernoulli Institute, University of Groningen, Nijenborgh 9, 9747AG,Groningen, Netherlands,
 \newline
 email: aenter@phys.rug.nl},\\ 
}

\maketitle

\begin{center}
{\bf Abstract:} 
\end{center}
 It is well-known that discrete-time finite-state Markov Chains, which are described by one-sided conditional probabilities which describe a dependence on  the past as only dependent on the present,  can also be described as one-dimensional Markov Fields, that is,  nearest-neighbor Gibbs measures for finite-spin models, which are described by two-sided conditional probabilities. In such Markov Fields the time interpretation of past and future is  being replaced by the space interpretation of an interior volume, surrounded by an exterior to the left and to the right. \\
 If we relax the Markov requirement to weak dependence, that is, continuous dependence, either on the past (generalising the Markov-Chain description) or on the external configuration (generalising the Markov-Field description), it turns out this equivalence breaks down, and neither class contains the other. In one direction this result has been known for a few years, in the opposite direction a counterexample was found recently. Our counterexample is based on the phenomenon of entropic repulsion in long-range Ising (or "Dyson") models.

\footnotesize

\vspace{7cm}



\normalsize
\section{Introduction}
It has been known since more than 40 years that finite-state discrete-time  Markov Chains are equivalent to Markov Fields (one-dimensional nearest-neighbour finite-spin Gibbs measures) \cite{HOG}, Ch 3. This result  was independently obtained by Brascamp and Spitzer.\\
In a Markov Chain the future is independent of the past, given the present; in a Markov Field the inside of a (finite) area is independent of the outside, given the border. If space is  $\Z$, one-dimensional, and time is discrete (thus also the one-dimensional  integer line $\Z$), the difference in description  is therefore between a one-sided (time-like) versus a two-sided (space-like) conditioning. But despite this, one obtains the same class of measures, as long as the conditioning is Markovian.\\
 A Markovian modeller displays a certain short-sightedness:\\
 For a Markov-Chain modeller, if one knows the present one controls the future ({\em ``All History is Bunk''} (Henry Ford)); and for a Markov-Field modeller, to rule an area it suffices to {\em ``Control the Borders''}. And in one dimension these are indeed the same. \\  
On the other hand, if we consider arbitrary stochastic processes, one-sided descriptions and two-sided descriptions can provide highly non-equivalent results.\\
A famous example thereof is the existence of measures which are ergodic and have a positive -- one-sided -- Kolmogorov-Sinai entropy,  (and have possibly even  a one-sided trivial tail), thus  being one-sided "stochastic",  despite having a full two-sided tail, therefore being two-sided deterministic  \cite{Gur,OW} and thus having zero two-sided entropy density.  We remark that the quantity which we here call a ``two-sided `` entropy, is a one-dimensional  example of  entropy-like quantities which can be defined on more general graphs; such quantities have been  also called  ``inner'' or ``conditional'' or ``lower'' or "erasure" entropies in the literature \cite{AvELMP,EV, FS,Temp,VW}.\\
These above examples, due  to Gurevich, and to Ornstein and Weiss,  provided non-equivalence of one-sided and two-sided entropies \cite{Temp}, one-sided and two-sided
tail  properties, \cite{HolSt} etc. We  notice that many quantities and properties from Ergodic Theory (Kolmogorov-Sinai entropy, K-property, Bernoulli property, Isomorphism as Dynamical Systems)  are defined in terms of one-sided objects or descriptions. \\

However, those examples lack continuity properties of their conditional probabilities. Such continuity properties in space  in fact characterise {\em Gibbs measures} \cite{Koz,Sul}. 

Thus the Gurevich and Ornstein-Weiss examples are excluded if one restricts oneself to the class of Gibbs measures.\\
In fact, it is known that one-sided and two-sided entropy densities for Gibbs measures (for absolutely summable potentials)  are identical \cite{Temp, AvELMP}, and that under some stronger conditions on the interaction decay, one-sided and two-sided tail properties of Gibbs measures are the same \cite{HolSt}.\\

For Gibbs measures in one dimension, it was open for a long time if one-sided and two-sided characterisations were equivalent.\\
 The measures which have one-sided continuous conditional probabilities are known as\\ $g$-measures (or ``chains with complete connections'' or `` chains of infinite order'', or ``random Markov chains''). They were introduced in the thirties and repeatedly rediscovered (under different names) \cite{DoeFor, Ruma, Harr,Keane, Kalik}. A few years ago, Fern\'andez,  Gallo and Maillard \cite{FGM} constructed a $g$-measure -- with one-sided continuous conditional probabilities-- which is not a Gibbs measure, as its two-sided conditional probabilities are not continuous.

Here we discuss our  \cite{ BEEL} recently finding  an opposite result, namely  that the Gibbs measures of the Dyson models --which have two-sided continuous conditional probabilities-- are not $g$-measures, as their one-sided conditional probabilities are not continuous. \\
Note that, in contrast to all the earlier  counterexamples of Gurevic, Ornstein-Weiss, and Fern\'andez-Gallo-Maillard, in our case the two-sided   behaviour is more ``regular'', more \\``stochastic'',  than the one-sided behaviour. 

\section{Background and Notation}
\subsection {Dyson Models}
Here we describe some  properties of one-dimensional long-range spin models, 
also known as Dyson models. \\
In his original work, Dyson \cite{Dys} considered an Ising spin system in one dimension (on $\Z$), with formal Hamiltonian given by 
\begin{equation}
H(\omega) = - \sum_{i>j} J(|i-j|)\omega_i\omega_j
\end{equation}
and $J(n) \geq 0$ for $n\in \mathbb{N}$ is of the form $J(n)=n^{-\alpha}$. \\

A conjecture due to  Kac and Thompson  \cite{Kac} had stated that there should be a phase transition for low enough temperatures if and only if $\alpha \in (1,2]$ (in zero magnetic field). Dyson proved a part of the Kac-Thompson conjecture, namely that for long-range models with interactions of the form $ J(n) = n^{-\alpha}$ with $\alpha \in (1,2)$, there is a phase transition at low temperatures.\\ 
Later different proofs were found, \cite{FILS,Joh,CFMP,ACCN} and also the case $\alpha=2$ was shown to have a transition \cite{FrSp}. 

In summary the following  holds: 

\begin{proposition}
\label{DyFrSp} 
\cite{Dys,FrSp,Rue72,HOG,FILS,ACCN,CFMP,LP,Joh}.
The Dyson model with polynomially decaying potential, for $1< \alpha \leq 2$, exhibits a {\em phase transition at low temperature}: 
$$ \exists \beta_c^D >0, \; {\rm such \; that} \; \beta > \beta_c^D \; \Longrightarrow \; \mu^- \neq \mu^+ \; {\rm and} \; \mathcal{G}(\gamma^D)=[\mu^-,\mu^+] $$
 where the extremal  measures $\mu^+$ and $\mu^-$ are translation-invariant. They have in particular opposite magnetisations   $\mu^+[\sigma_{0}]=-\mu^-[\sigma_{0}]=M_0(\beta, \alpha)>0$ at low temperatures. Moreover, the Dyson  model in a non-zero homogeneous field $h$ has a unique Gibbs measure at all temperatures. 
\end{proposition}

It is well-known that  there is no phase transition for $J(n)$ being of finite range, and neither for $J(n)=n^{-\alpha}$ with $\alpha > 2$. \\

\begin{remark}
The case of $\alpha=2$ is more complicated to analyse, and richer in its behaviour, than the other ones. There exists a hybrid transition (the "Thouless effect"), as the magnetisation is discontinuous while the energy density is continuous at the transition point. Moreover, there is second transition below this transition temperature. In the intermediate phase there is a positive magnetisation with non-summable covariance, while at very low temperatures the covariance decays at the same rate as the interaction, which is summable. For these results, see  \cite{ACCN, I,IN}.
\end{remark}
Here we will make use of the approach of \cite{CFMP}, which has been extended to a number of other situations (Dyson models in random fields \cite{COP}, interfaces \cite{CMPR}, phase separation \cite{CMP17},  inhomogeneous decaying fields \cite{BEEKR}, etc). The disadvantage of this approach is that it works only at very low temperatures, as it is perturbative, and it works only for a reduced set of $\alpha$-values, $\alpha^* < \alpha <  2$,  with ${\alpha}^{*} = 3 - \frac{\ln 3}{ \ln 2}$. The advantage, however,  compared to other proofs,  is that translation invariance does not play that much of a role. \\
The main idea of the approach of \cite {CFMP}, which was introduced in the $\alpha=2$ case by Fr\"ohlich and Spencer in \cite{FrSp}, is to construct a kind of triangular contours for which a Peierls-type contour argument can be obtained.  The energy of a contour of length $L$ has an energy cost associated to it of order $O(L^{2- \alpha})$, (and of order $O(\ln L)$ when $\alpha=2$).\\

\noindent 
There has been substantial interest in the Dyson model over the years.\\
Varying the decay parameter $\alpha$ plays a similar role as varying the dimension in short-range models. This can be done in a continuous manner, so one obtains  analogues of well-defined models in continuously varying non-integer dimensions. This is one major reason why these models have attracted a lot of attention in the study of phase transitions and critical behaviour (see e.g. \cite{CFMP} and references therein).\\
For some  recent results  for these long-range Dyson  models with polynomially decaying interactions, see  \cite{LP, ELN, BEEKR, CMP17, EKRS, CELR}.

\medskip

\subsection{Specifications and Measures}

We refer to \cite{Bov, HOG,EFS,Fer,FV, Rue} for more  general treatments of the 
Gibbs formalism.\\ 
Dyson models  are  special, as they are {\em ferromagnetic} Ising models 
with   long-range {\em pair} interactions in {\em one dimension}.\\ 
 We consider  these models  as belonging to  a  more general class of lattice (spin) models with Gibbs measures on infinite-volume product configuration spaces $(\Omega,\mathcal{F},\rho)=(E^{{\Z}^d},\mathcal{E}^{{\otimes {\Z}^d}},\mu_o^{{\otimes {\Z}^d}})$. In our case $d=1$, and  the single-site state space is  the Ising space 
$E=\{-1,+1\}$,
with the a priori counting 
measure $\mu_0=\frac{1}{2} \delta_{-1} + \frac{1}{2} \delta_{+1}$. We denote by 
$\mathcal{S}$ the set of the finite subsets of $\Z$ and, for any $\La \in \s$, 
write $(\Om_\La,\mathcal{F}_\La,\rho_\La)$ for the finite-volume configuration 
space $(E^\La,\mathcal{E}^{\otimes \La},\mu_o^{\otimes \La})$.  
We also will consider only translation-invariant models.

Microscopic states or configurations, denoted by $\si,\om, \eta, \tau,\;$ etc., are  elements of  $\Omega$, 
equipped with the product topology of the discrete topology on $E$, for which these configurations are close when they coincide on large finite regions $\Lambda$ (the larger the region where they are equal, the closer the configurations are). 


We denote by $C(\Om)$ the set of continuous  (quasilocal) functions on $\Om$,  characterized by

\be \label{qlocfu} 
f \in C(\Omega) \; \Longleftrightarrow \; \lim_{\Lambda \uparrow \Z} \sup_{\sigma,\omega:\sigma_\Lambda=\omega_\Lambda} \mid f(\omega) - f(\sigma) \mid = 0.
\ee

The fact that we consider {\em ferromagnetic pair} interactions provides us with an extra tool:\\
We can make use of FKG inequalities.
Monotonicity for functions and measures concerns the natural partial (FKG) \cite{FKG} order "$\leq$ ", which we have on our Ising spin systems : $\sigma \leq \omega$ if and only if 
$\sigma_i \leq \omega_i$ for  all  $i \in \Z$. Its maximal and minimal elements 
are the configurations $+$ and $-$, and this order extends to functions: 
$f:\Omega \longrightarrow \mathbb{R}$ is called {\em monotone increasing}  when 
$\sigma \leq \omega$ implies $f(\sigma) \leq f(\omega)$. For measures,  we write $\mu \leq \nu$ if and only
if $\mu[f] \leq \nu[f]$ for all $f$ monotone increasing\footnote{We denote 
$\mu[f]$ for the expectation $\E_\mu[f]$ under a measure $\mu$.}.
FKG arguments are based on the observation that increasing the interaction, or an external field, will FKG-increase the associated Gibbs measure. 
\\

Macroscopic states are  represented by probability measures  
on  $(\Omega,\mathcal{F},\rho)$. 

Following Dobrushin, Lanford and Ruelle, DLR or Gibbs measures are defined in terms of consistent systems of (regular versions of) finite-volume conditional probabilities, of finite-volume configurations with prescribed boundary conditions outside of those volumes \cite{Dob1,LaR}.\\
Such a family of (everywhere, rather than almost everywhere, defined, as one has not yet a measure,  which could determine that something is  almost sure with respect to it) conditional probabilities is called a {\em specification}.\\
A measure for which a version of its  conditional probabilities provides  those of the specification is said to satisfy the DLR conditions for that specification.  
The specifications of interest in the theory of lattice systems usually satisfy
a finite-energy condition. This says that no configuration in a local volume is excluded (has probability zero), uniformly in the boundary conditions on which one conditions. Moreover, the condition of continuity (or quasilocality) is required. This says that all conditional probabilities are continuous (quasilocal) functions of the boundary conditions.\\

\begin{remark}
In contrast to Kolmogorov's theorem, which says that a system of consistent marginal probabilities defines precisely one probability measure, for finite-spin specifications the number of measures satisying the DLR conditions for that specification can be either one or more. The latter situation sometimes is taken as the definition of a {\bf Phase Transition}. 
\end{remark} 

A {\it measure} is said to be  quasilocal when it is specified by a quasilocal specification. 

A particularly important approach to  quasilocal measures consists in considering  the {\em Gibbs measures} with (formal) Hamiltonian $H$ defined  via a potential (or interaction) $\Phi$, a family $\Phi=(\Phi_A)_{A \in \s}$ of local functions $\Phi_A \in \mathcal{F}_A$.  The contributions of spins in finite sets $A$ to the total energy define the {\em finite-volume  Hamiltonians with free boundary conditions} 
\be \label{Ham}
\forall \Lambda \in \s,\; H_\Lambda(\omega)=\sum_{A \subset \Lambda} \Phi_A(\omega),\; \forall \omega \in \Omega.
\ee
To define Gibbs measures, we require for $\Phi$ that it is  {\em Uniformly Absolutely Summable} (UAS), i.e. 
that $\sum_{A \ni i} \sup_\omega |\Phi_A(\omega)| < \infty, \forall i \in \Z$.
One then can give sense to the  {\em Hamiltonian at \\volume $\Lambda \in \s$ with boundary condition $\omega$} defined for all $\sigma,\omega \in \Om$ as\\ $H_\Lambda^\Phi(\sigma | \omega) := \sum_{A \cap \Lambda \neq \emptyset} \Phi_A(\sigma_\Lambda \omega_{\Lambda^c}) (< \infty)$.
The {\em Gibbs specification at inverse temperature $\beta>0$} is then defined by
\be \label{Gibbspe}
\gamma_\Lambda^{\beta \Phi}(\sigma \mid \omega)=\frac{1}{Z^{\beta \Phi}_\Lambda(\omega)} \; e^{-\beta H_\Lambda^\Phi(\sigma | \omega)} (\rho_\Lambda\otimes \delta_{\omega_{\Lambda^c}}) (d \sigma)
\ee
where the partition function $Z_\Lambda^{\beta \Phi}(\omega)$ is a normalizing constant. Due to the
UAS condition, these specifications are quasilocal. It turns out that the converse is also true up to a non-nullness condition\footnote{expressing that $\forall \Lambda \in \s,\; \forall A \in \mathcal{F}_\Lambda$, $\rho(A)>0$ implies that $\gamma_\Lambda (A | \omega) >0$ for any  $\omega \in \Om$.} (see e.g. \cite{HOG, Fer, Koz, Sul, ALN2}) and one can take :
\begin{definition}[Gibbs measures]:\\
$\mu \in \mathcal{M}_1^+$ is a Gibbs measure iff $\mu \in \mathcal{G}(\gamma)$, that is, the conditional probabilities of $\mu$ - at least  a version thereof- are those given by $\gamma$  where $\gamma$ is a non-null and quasilocal specification.
\end{definition}

\be \label{esscont}
\lim_{\Delta \uparrow \mathbb{Z}} \sup_{\omega^1,\omega^2 \in \Omega}  \Big| \mu \big[f |\mathcal{F}_{\Lambda^c} \big](\omega_\Delta \omega^1_{\Delta^c}) - \mu \big[f |\mathcal{F}_{\Lambda^c} \big](\omega_\Delta\omega^2_{\Delta^c})\Big|=0
\ee
Thus, for Gibbs measures the conditional probabilities always have continuous versions, or equivalently
 there is no point of essential discontinuity. Points of essential discontinuity are  configurations which are points of discontinuity for ALL versions of the conditional probability. In particular one cannot make conditional probabilities continuous by redefining them on a measure-zero set if such points exist. In the generalized Gibbsian framework, one also says that such a configuration is a {\em bad configuration} for the considered measure, see e.g. \cite{ALN2}.
The existence of such bad configurations implies non-Gibbsianness of the associated measures.\\

\begin{remark}
 If the interaction is of finite range (a Markov Field), or sufficiently fast decaying, uniqueness of the Gibbs measure holds; indeed  no phase transition is expected in one dimension in considerable generality. But the Dyson model gives a counterexample if just UAS is required. 
\end{remark}

\begin{remark}
In fact it is enough to know the single-site conditional probabilities -- the single-site specification --, all other conditional probabilities can be obtained from those. 
\end{remark}



Although the extension of the DLR equation to infinite sets is direct in case of uniqueness of the DLR-measure for a given  specification \cite{FP, Foll, Gold2},  it can be more problematic otherwise: it is valid for finite sets only and  measurability problems might arise in case of phase transitions  when one wants to extend them to infinite sets. Nevertheless, beyond the uniqueness case, such an extension was made possible by Fern\'andez and Pfister  \cite {FP} in the case of attractive models, that is models satisfying FKG properties.\\ 
 As we will make  use of it, we describe it now in our particular case. 
The concept they introduced is that of  a {\em global specification}.

A  global specification is a set of consistent conditional probabilities where one considers probabilities of sets which have their supports not only in finite sets, but in more general sets $S \in \Z$, which can be infinite, possibly  with infinite complements. The existence of such a global specification can be invoked to derive the existence of conditional probabilities of sets in  $E^S$, and the possibility of conditioning  on 
configurations in $E^{S^c}$.


Note, by considering $S=\mathbb{Z}$, that the set of measures a version of whose conditional probabilities is given by a global specification  contains at most one element. \\

The case we will be most interested in is the situation where we condition on only one half-line. This leads us to the concept of $g$-measures.\\ The formalism of $g$-measures can be  developed in a parallel manner to the Gibbs formalism, but only using one-sided objects (conditional probabilities, specifications, etc.). \\
We will call a measure a $g$-measure, once the future depends in a continuous manner on the past.

\begin{definition}
Let $\mu$ be a  measure on $\Omega ={E}^\Z$. We will call $\mu$ a $g$-measure for the function $g$ if the conditional probability for the next symbol being $a$, $\mu(x_0 =a| \{\omega_i \}_{i< 0}) = g(\omega_{{\Z}^{-}}a)$, depends in a continuous manner on the past, that is,  $g$ is continuous function on ${E}^{\Z^{-}}$.  

\end{definition}  

For further description and background, we refer to \cite{BFV}\\
 
\begin{remark}
 We can also obtain the equivalent of a specification, a ``LIS'' (Left Interval Specification) for which the measure is a $g$-measure, analogously to the Gibbs measure definition \cite{FM1, FM2}. This is based on the observation that one can build general conditional probabilities from single-site ones.
\end{remark}

\begin{remark}
 Although in the Markov-Chain set-up (which is a simple and well-known example of a $g$-measure) uniqueness holds, in the general $g$-measure case, similarly to the Gibbsian set-up, under only  the  condition of continuity phase transitions are possible \cite{BK,BHS}. However, although under appropriate uniqueness conditions (uniform boundedness of boundary energies,  or Dobrushin uniqueness, for example) $g$-measures and Gibbs measures turn out to be the same objects (\cite{BFV,FM1,FM2}, no general equivalence seems to apply in a more general setting. Thus the connection between the different classes of phase transitions possible is mostly unknown (if there  even is one). 
\end{remark}

\section{From interface localisation to entropic repulsion,\\ and from there to the lack of the g-measure property:\\ One Ingredient and three Observations }

I) {\bf The Main Ingredient:\\ Mesoscopic Interface Localisation}: \\

In \cite{CMPR} the main result describes the {\em mesoscopic} localisation of an ``interface point'' at low temperatures, for a Dyson model with Dobrushin boundary conditions (minus to the left, plus to the right).\\
 If one considers an interval $[-L,+L]$ with such Dobrushin boundary conditions, with probability close to 1 the interface point is near the center, that is, with large probability it is not more than 
$O(L^{\frac{\alpha}{2}})$
away from  the center with a (Gaussian) probability distribution; moreover the probability for the interface point  being at larger distances than $\varepsilon L$ from the origin  is bounded by $ O(L \exp -L^{2 - \alpha})$. The system is in a minus-like phase with strictly negative magnetisation left of the interface point, and in a plus-like phase with positive magnetisation to the right of the interface point.\\ 
Although there are some extra  conditions required (on the decay power, and the strength of the nearest-neighbour term), due to the use of the triangular contours of \cite{CFMP}, it seems that one can rid of those with enough effort, see \cite{LP,BEEKR}. Moreover, for a counterexample, of course we need not strive for the greatest generality, so we assume that all necessary conditions are satisfied (and there are enough situations known where this happens).\\

II) {\bf First Observation:\\ From Interface Localisation To Entropic Repulsion.}\\
  
It is a simple observation that if we change all spins at sites left of $-N$ from minus to plus, with $N$ chosen large enough so that $L \times  N^{1- \alpha}$ is small ( in particular, then $N$ is much larger than $L$), this does not change the probability distribution in the interval $[-L, L]$ by that much. If the energy differences are small, so are the differences in (conditional) Gibbsian probabilities.  \\
If we move the right border further to the right than $+L$, the interface point can only move to the right, due to an FKG argument. This implies that if we take the plus measure ${\mu}^{+}$ and condition on a large $O(N)$ interval to be minus, typically at a distance at least of size $L$ one finds oneself in a minus-like phase. The interpretation of this is that a ``hard'' ({\em frozen}) minus-interval pushes an interface away, due to an entropic repulsion mechanism. The minus-phase can be seen  in terms of a wetting phenomenon, as a ``wet region''.\\
It should be noted that the effect requires positive, non-zero temperature. \\
Moreover for the case $\alpha =2$  there is  {\em no mesoscopic} localisation, but the position of the interface point has {\em macroscopic} fluctuations. Thus in that case, our proof breaks down.

III) {\bf Second Observation:\\ Decoupling A Not Too Large Interval Does Not Shift The Interface.}\\
The position of the interface point does not change much if, next to the frozen minus interval, we decouple all the sites in a large but not too large interval to the right of it (that is at distance less  than $L_0$, with $L_0(L)$ much less than $L$, but possibly diverging with growing $L$).\\ 
This follows again from the contour analysis of \cite{CMPR}. Moving the interface point by a macroscopic distance (proportional to $L$) costs an energy which diverges with $L$. If the total energy of an interval of size $L_0$ is less than that, then the interface point won't move on a macroscopic (O(L)) scale. 

IV) {\bf Third Observation:\\ Changing A  Decoupled  Interval To An Alternating One Costs Finite Energy.} 

If we choose an alternating configuration in the interval of size $L_0$, and restore its\\ interaction with its complement, the interaction energy with its exterior is uniformly bounded.\\
Namely, the sum below satisfies: \\
$\sum_{i=1 ... L_0}\sum_{k > L_0} |k-i|^{-\alpha}(-1)^{i} {\omega}_k = \\
\sum_{i=0 ... L_0} \sum_{k > L_0} O(|k-i|^{-\alpha} - {|k+1-i|}^{- \alpha}) <\\
\sum_{i < L_0} \sum_{k > L_0} O(|k-i|^{- (\alpha +1)} < \infty$.\\

It is known that finite-energy perturbations will cause only relatively small, essentially microscopic, changes \cite{BLP}. In particular, shifting the interface point costs a finite energy only if the shift is over a finite distance.\\ 

Thus a large  alternating interval, preceded by a (VERY) large frozen minus interval configuration  again is succeeded by a minus phase, while, when it is preceded by a plus interval, it is succeeded by a plus-phase interval. But this dependence on the presence of  a frozen plus or minus interval far (of order $L_0$) to the left (= in the past), violates the continuity condition which is required for ${\mu}^{+}$ to be a  $g$-measure. As our  measure was defined to be a Gibbs  measure for the Dyson interaction, it automatically has two-sided continuity; thus we have obtained our counterexample.  

We can therefore conclude \cite{BEEL}:

\begin{theorem}
The low-temperature Gibbs measures of Dyson models cannot be written as $g$-measures for a continuous $g$-function. Therefore, of the class of Gibbs measures for quasilocal specifications  and the class of $g$-measures with continuous $g$-functions, neither of the two classes contains the other one.    
\end{theorem}

\section{Conclusion, Final Remarks, Higher Dimensions}

We have shown that one-sided and two-sided continuity of conditional probabilities not only are not equivalent, but that neither of the two continuity conditions  implies the other one.\\ 
In other words, controlling your borders and controlling the future are not the same things, except if you are a short-sighted Markovian.\\  
Although the result in \cite{BEEL} was proven under some restrictions on $\alpha$, and the proof also requires the presence of a large enough nearest-neighbour interaction, in view of the results of \cite{LP,BEEKR} these conditions can presumably be removed.\\
However, the situation for $\alpha=2$ remains unclear, as the main idea of our proof (mesoscopic localisation of the interface point) fails. It is not clear to us what to expect in that case.

In higher dimensions, the Markov-Field property occurs as a ``Local'' Markov property, whereas other properties, such as the Global Markov property or being a 
(tree-indexed) Markov Chain (a ``splitting Gibbs measure'') on tree graphs, play a role which looks more like the Markov-Chain property. It is known, however, that in contrast to the one-dimensional situation  these properties do not follow from the Local Markov properties \cite{Foll, Gold2,Isr,Roz, HOG}.\\ 
Moreover, if one tries to compare such non-local Markov properties with nonlocal  continuity properties by for example considering continuity properties as a function of the {\em lexicographic past}, (as the analogue of the one-dimensional ordinary past), it is not difficult to see that the low-temperature plus-phase of the Ising model in $d=2$, for example, although it is actually known to be well-behaved enough that it even satisfies the Global Markov Property, has conditional probabilities which display an essential point of discontinuity. So even a Global Markov Field can have conditional probabilities which are discontinuous as a function of the (lexicographic) past. The proof of this statement is very close to that of the non-Gibbsianness of the Schonmann projection \cite{Sch}  (the marginal of the low-temperature two-dimensional Gibbs measure on the configurations a line $\{\Z,0\}$) as given in \cite{EFS}, and further studied in \cite{FP,BC} for example; only we now replace the line $\{\Z , 0 \}$ surrounding the origin by two half-lines $ \{{\Z}^-,0 \} $ and $ \{ {\Z}^+ \cup  0,-1 \}$ in the lexicographic past of the origin, then the proof goes through more or less literally. The wetting phenomenon which is responsible is identical: there is an entropic repulsion from a frozen interval into the ``future'' direction producing a wet droplet, and having two intervals left and right which are large enough, causes the two wet droplets to merge. Under conditions of strong uniqueness (high-temperature Dobrushin uniqueness, or Dobrushin-Shlosman conditions e.g.), continuity of the magnetisation in the origin as a function of the lexicographic past  configurations holds, however. (I thank Brian Marcus and Siamak Taati for asking me this question and discussions on this issue).

\newpage

{\bf Acknowledgments:} 
I dedicate this paper with pleasure to Anton Bovier, on the occasion of his 60th birthday. Anton has been a reliable and stimulating guide for me in matters literary, culinary and scientific, for more than 30 years. His many-sided personality and friendship have enriched my life in  many respects. I wish him many more years, and look forward to have a share therein.

I thank my collaborators and discussion partners on Dyson models, long-range models and $g$-measures  for the pleasure of these collaborations and for all I have learned from them. In alphabetical order I owe: Stein Bethuelsen, Rodrigo Bissacot, Diana Conache, Loren Coquille, Eric Endo, Roberto Fern\'andez, Frank den Hollander, Bruno Kimura, Arnaud Le Ny, Brian Marcus, Pierre Picco, Wioletta Ruszel, Cristian Spitoni, Siamak Taati and Evgeny Verbitskiy.\\ 
I thank Eric,  Siamak and Arnaud for helpful advice on the manuscript.\\
I thank Louis-Pierre Arguin, Veronique Gayrard, Nicola Kistler and Irina Kourkova for inviting me to be present at the CIRM meeting to celebrate Anton, and also to contribute to this volume.  


\end{document}